\documentclass[a4paper,11pt]{spieman}  
\usepackage{amsmath,amsfonts,amssymb}
\usepackage{graphicx}
\usepackage{setspace}
\usepackage{tocloft}

\title{e-ASTROGAM mission: a major step forward for gamma-ray polarimetry}

\author[a,*]{V. Tatischeff}
\author[b,c]{A. De Angelis}
\author[d]{C. Gouiff\`es}
\author[e]{L. Hanlon}
\author[f]{P. Laurent}
\author[g]{G. M. Madejski}
\author[h,i,j]{M. Tavani}
\author[e]{A. Ulyanov}
\author[k]{on behalf of the e-ASTROGAM Collaboration}
\affil[a]{Centre de Sciences Nucl\'eaires et de Sciences de la Mati\`ere, IN2P3-CNRS/Univ. Paris-Sud, Universit\'e Paris-Saclay, F-91405 Orsay Campus, France}
\affil[b]{Universit\`a di Padova and INFN, via Marzolo 8, I-35131 Padova, Italy}
\affil[c]{LIP/IST, Av. Elias Garcia 14, 1000 Lisboa, Portugal}
\affil[d]{Laboratoire AIM, UMR 7158, CEA/DRF, CNRS, Universit\'e Paris Diderot, IRFU/SAp, F-91191 Gif-sur-Yvette, France} 
\affil[e]{School of Physics, University College Dublin, Belfield, Dublin 4, Ireland}
\affil[f]{Laboratoire APC, UMR 7164, CEA/DRF CNRS, Universit\'e Paris Diderot, Paris, France}
\affil[g]{Kavli Institute for Particle Astrophysics and Cosmology, Department of Physics and SLAC National Accelerator Laboratory,
Stanford University, Stanford, CA 94305, USA}
\affil[h]{INAF-IAPS, Via del Fosso del Cavaliere 100, I-00133 Roma, Italy}
\affil[i]{INFN Roma Tor Vergata, Via della Ricerca Scientifica 1, I-00133 Roma, Italy}
\affil[j]{Dipertimento di Fisica, Università di Roma "Tor Vergata", Via Orazio Raimondo 18, I-00173 Roma, Italy}
\affil[k]{See \url{http://eastrogam.iaps.inaf.it/}}

\cftpagenumbersoff{figure}
\cftpagenumbersoff{table} 
\def\gsim{\lower.5ex\hbox{$\; \buildrel > \over \sim \;$}}
\def\lsim{\lower.5ex\hbox{$\; \buildrel < \over \sim \;$}}
\begin{document} 
\maketitle

\begin{abstract}
e-ASTROGAM is a  gamma-ray space mission proposed for the fifth Medium-size mission (M5) of the European Space Agency. It is dedicated to the study of the non-thermal Universe in the photon energy range from $\sim$0.15~MeV to 3 GeV with unprecedented sensitivity, angular and energy resolution, together with a groundbreaking capability for gamma-ray polarimetric measurements over its entire bandwidth. We discuss here the main polarization results expected at low energies, between 150 keV and 5 MeV, using Compton interactions in the e-ASTROGAM instrument, from observations of active galactic nuclei,  gamma-ray bursts, microquasars, and the Crab pulsar and nebula. The anticipated performance of the proposed observatory for polarimetry is illustrated by simulations of the polarization signals expected from various sources. We show that polarimetric analyses with e-ASTROGAM should provide definitive insight into the geometry, magnetization and content of the high-energy plasmas found in the emitting sources, as well as on the processes of radiation of these plasmas. 
\end{abstract}


{\noindent \footnotesize\textbf{*} \linkable{Vincent.Tatischeff@csnsm.in2p3.fr} }

\begin{spacing}{1}   %

\section{Introduction}
\label{sect:intro}  

Gamma-ray polarimetry can be a powerful diagnostic of the high-energy physics at work in cosmic sources emitting collimated outflows and jets (e.g. blazars, gamma-ray bursts, X-ray binaries) or with strong magnetic field (e.g. pulsars, magnetars). In addition, measuring the linear polarization of distant (i.e. cosmological) gamma-ray sources (blazars, gamma-ray bursts) is a means of studying fundamental questions of physics related to Lorentz invariance violation (Ref.~\citenum{lau11b}). However, very few polarization results have been obtained in high-energy astronomy so far, due to the limited sensitivity of existing instruments to this observable. 

The e-ASTROGAM mission concept aims to fill the gap in our knowledge of astronomy in the medium-energy (∼0.3–-100 MeV) gamma-ray domain by increasing the number of known sources in this field by more than an order of magnitude and providing polarization information for many of them.  Between 3000 and 4000 sources are expected to be detected during the first three years of mission operation. The e-ASTROGAM gamma-ray instrument inherits from its predecessors such as {\it AGILE}\cite{tav09} and {\it Fermi}\cite{atw09}, as well as from the MEGA (Medium Energy Gamma-ray Astronomy) prototype\cite{kan05}, but takes full advantage of recent progress in silicon detectors and readout microelectronics to achieve excellent spectral and spatial resolution by measuring the energy and 3D position of each interaction within the detectors. The e-ASTROGAM mission concept is presented at length in Ref.~\citenum{eas}. Here, we first give an overview of the proposed observatory (Sect.~\ref{sect:ea}) and then outline the breakthrough capability of the e-ASTROGAM telescope for gamma-ray polarimetric observations of some of the main targets of the mission: active galactic nuclei (AGN), gamma-ray bursts (GRBs), the Crab pulsar/nebula system, and microquasars (Sect.~\ref{sect:polarimetry}).

\section{The e-ASTROGAM observatory}
\label{sect:ea}  

\begin{figure}
\centering
\includegraphics[width=0.75\textwidth]{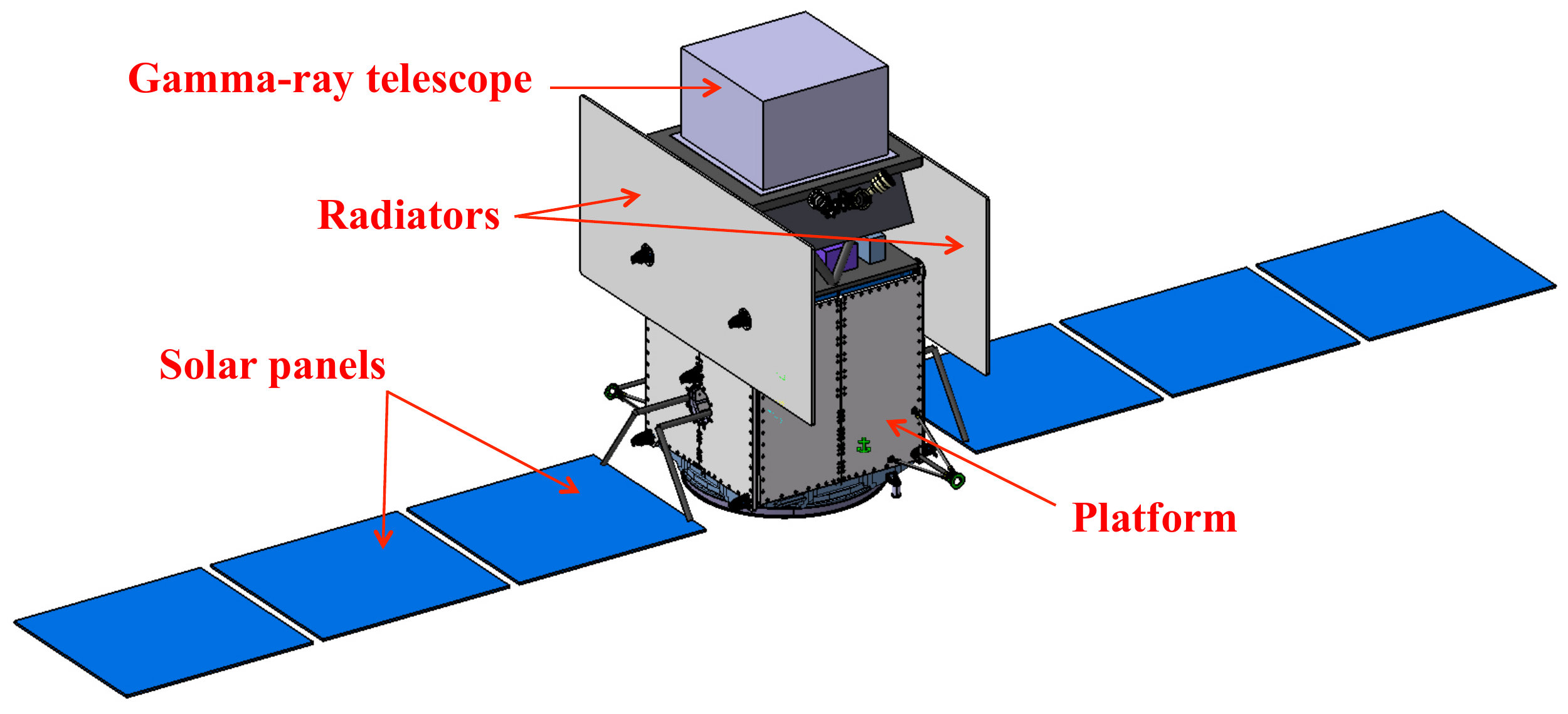}
\caption{e-ASTROGAM spacecraft with solar panels deployed.}
\label{fig:deployed}
\end{figure}

The payload of the e-ASTROGAM satellite (Figure~\ref{fig:deployed}) consists of a single gamma-ray telescope operating over more than four orders of magnitude in energy (from about 150 keV to 3 GeV) by the joint detection of photons in both the Compton (0.15 -- 30 MeV) and pair ($> 10$~MeV) energy ranges. It is attached to a mechanical structure at a distance of about 90~cm from the top of the spacecraft platform, the space between the payload and the platform being used to: (i) host a time-of-flight (ToF) unit designed to discriminate between particles coming out from the telescope and those entering the instrument from below; (ii) host several units of the payload (the back-end electronics modules, the data handling unit, and the power supply unit) and (iii) accommodate two fixed radiators of the thermal control system, each of 5.8~m$^2$ area (Figure~\ref{fig:deployed}). This design has the advantage of significantly reducing the instrument background due to prompt and delayed gamma-ray emissions from fast particle reactions with the platform materials. 

\begin{figure}
\begin{center}
\begin{tabular}{c}
\begin{minipage}{0.42\linewidth}
\includegraphics[scale=0.3]{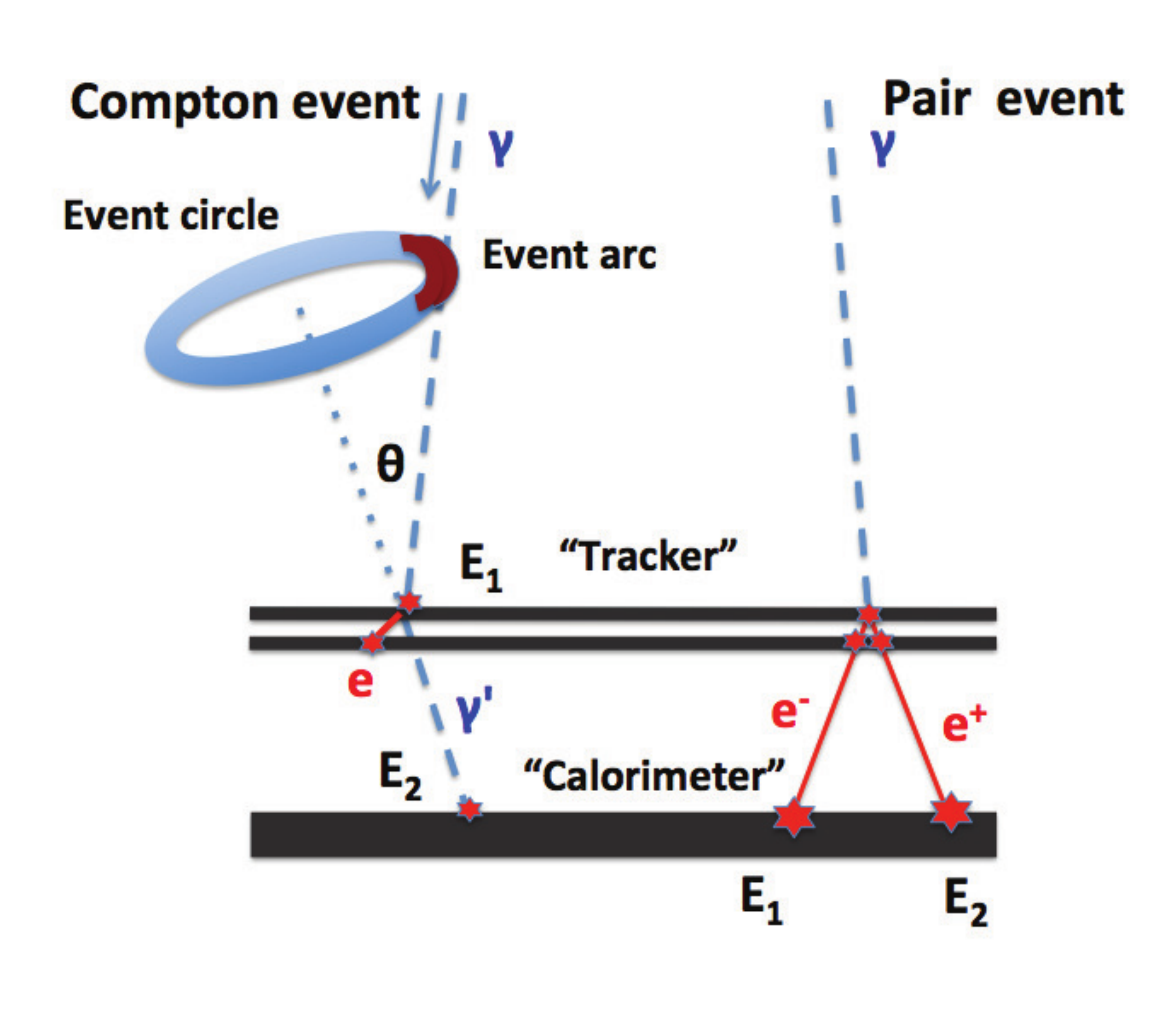}
\end{minipage}
\begin{minipage}{0.58\linewidth}
\includegraphics[scale=0.22]{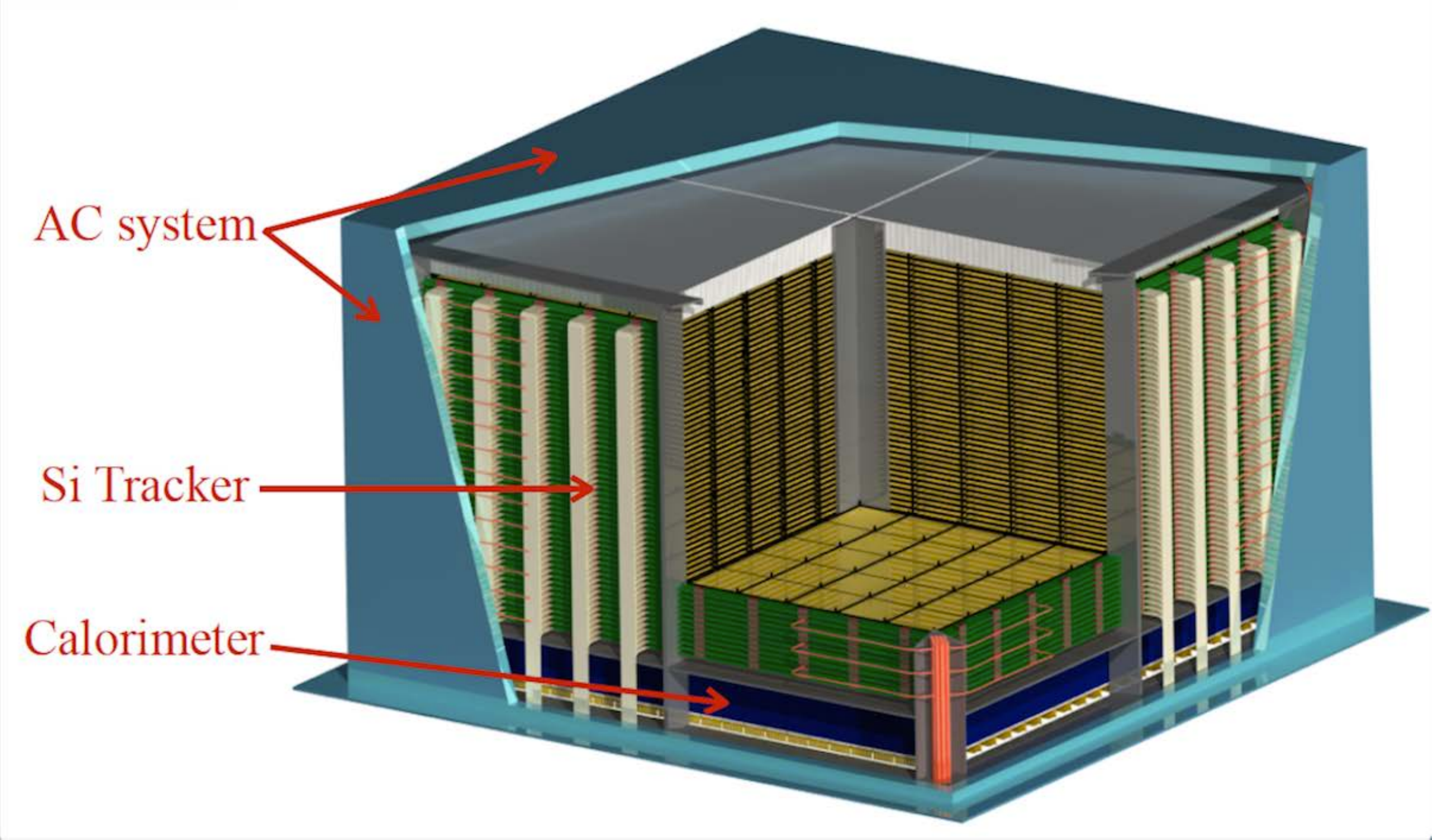}
\end{minipage}
\\
(a) \hspace{8.2cm} (b)
\end{tabular}
\end{center}
\caption 
{ \label{fig:payload}
(a) Representative  topologies for a Compton event and  for a pair event. Photon tracks are shown in pale blue, dashed, and electron and/or positron tracks are in red, solid.  (b) Overview of the e-ASTROGAM payload.} 
\end{figure}

The e-ASTROGAM telescope is made up of three detection systems (Figure~\ref{fig:payload}): a silicon Tracker in which the cosmic gamma-rays undergo a Compton scattering or a pair conversion (see Figure~\ref{fig:payload}a); a Calorimeter to absorb and measure the energy of the secondary particles and an anticoincidence (AC) system to veto the prompt-reaction background induced by charged particles. The telescope has a size of 120$\times$120$\times$78 cm$^3$ and a mass of 1.2~tons (including maturity margins plus an additional margin of 20\% at system level).

The Si Tracker comprises 5600 double-sided strip detectors (DSSDs) arranged in 56 layers. It is divided in four units of 5$\times$5 DSSDs, the detectors being wire bonded strip to strip to form 2-D ladders. Each DSSD has a geometric area of 9.5$\times$9.5 cm$^2$, a thickness of 500~$\mu$m, and a strip pitch of 240~$\mu$m. The total detection area amounts to 9025 cm$^2$. Such a stacking of relatively thin detectors enables efficient tracking of the electrons and positrons produced by pair conversion, and of the recoil electrons produced by Compton scattering. The DSSD signals are read out by 860,160 independent, ultra low-noise and low-power electronics channels with self-triggering capability.

The Calorimeter is a pixelated detector made of a high-$Z$ scintillation material -- Thallium activated Cesium Iodide -- for efficient absorption of Compton scattered gamma-rays and electron-positron pairs. It consists of an array of 33,856 parallelepiped bars of CsI(Tl) of 8~cm length and 5$\times$5~mm$^2$ cross section, read out by silicon drift detectors (SDDs) at both ends, arranged in an array of 529 ($=23 \times 23$) elementary modules each containing 64 crystals. The depth of interaction in each crystal is measured from the difference of recorded scintillation signals at both ends. Accurately measuring the 3D position and deposited energy of each interaction is essential for a proper reconstruction of the Compton events. The Calorimeter thickness -- 8 cm of CsI(Tl) -- makes it a 4.3 radiation-length detector having an absorption probability of a 1-MeV photon on-axis of 88\%.

The third main detector of the e-ASTROGAM payload consists of an Anticoincidence system composed of two main parts: (1) a standard Anticoincidence, named Upper-AC, made of segmented panels of plastic scintillators covering the top and four lateral sides of the instrument, requiring a total active area of about 5.2~m$^2$, and (2) a Time of Flight (ToF) system, aimed at rejecting the particle background produced by the platform. The Upper-AC detector is segmented in 33 plastic tiles (6 tiles per lateral side and 9 tiles for the top) coupled to silicon photomultipliers (SiPM) by optical fibers. The bottom side of the instrument is protected by the ToF unit, which is composed of two plastic scintillator layers separated by 50 cm, read out by SiPMs connected to Time Digital Converters. The required timing resolution is 300 ps. 

For best environmental conditions, the e-ASTROGAM satellite should be launched into a quasi-equatorial (inclination $i < 2.5^\circ$) low-Earth orbit (LEO) at a typical altitude of 550~--~600~km. The background environment in such an orbit is now well-known thanks to the Beppo-SAX\cite{cam14} and {\it AGILE}\cite{tav09} missions. In addition, such a LEO is practically unaffected by precipitating particles originating from solar flares, a virtue for background rejection. 

Extensive simulations of the instrument performance using state-of-art numerical tools\cite{zog06,bul12} and a detailed numerical mass model of the satellite together with a thorough model for the background environment have shown that e-ASTROGAM will achieve\cite{tat16}: 
\begin{itemize}
\item Broad energy coverage ($\sim$0.15 MeV to 3 GeV), with nearly two orders of magnitude improvement of the continuum sensitivity in the range 0.3 -- 100 MeV compared to previous missions;
\item Excellent sensitivity for the detection of key gamma-ray lines e.g. sensitivity for the 847~keV line from thermonuclear supernovae 70 times better than that of the {\it INTEGRAL} spectrometer (SPI);
\item Unprecedented angular resolution both in the MeV domain and above a few hundreds of MeV  i.e. improving the angular resolution of the COMPTEL telescope on board the {\it Compton Gamma Ray Observatory} ({\it CGRO}) and that of the {\it Fermi}/Large Area Telescope (LAT) instrument by a factor of $\sim$4 at 5 MeV and 1~GeV, respectively (e.g. the e-ASTROGAM Point Spread Function (68\% containment radius) at 1 GeV is 9').
\item Large field of view ($>$ 2.5 sr), ideal to detect transient Galactic and extragalactic sources, such as X-ray binaries and gamma-ray bursts;
\item Timing accuracy of 1~$\mu$s (at 3$\sigma$), ideal to study the physics of magnetars and rotation-powered pulsars, as well as the properties of terrestrial gamma-ray flashes;
\item Pioneering polarimetric capability for both steady and transient sources, as illustrated in the next Section. 
\end{itemize}

\section{Polarimetry with e-ASTROGAM}
\label{sect:polarimetry}  

e-ASTROGAM will be sensitive to the linear polarization of incident gamma-rays over its entire bandwidth. In the Compton range, the polarization signature is reflected in the probability distribution of the azimuthal scatter angle. In the pair production domain, the polarization information is given by the distribution of azimuthal orientation of the electron-positron plane. e-ASTROGAM will have a breakthrough capacity for gamma-ray polarimetry thanks to the fine 3D position resolution of both the Si Tracker and the Calorimeter, as well as the light mechanical structure of the Tracker, which is devoid of any heavy absorber in the detection volume. 

The measurement of polarization in the pair creation range, using the azimuthal orientation of the electron-positron plane, is complex and a precise evaluation of the unfolding procedures and performance requires accurate simulation and testing \cite{dbpol}. In the following, we focus on the e-ASTROGAM performance for polarimetry in the Compton domain. We discuss in particular the polarimetric capability of e-ASTROGAM for the study of active galactic nuclei (AGN), gamma-ray bursts (GRBs), the Crab pulsar and nebula, as well as microquasars. e-ASTROGAM will explore for the first time the polarimetric properties of celestial sources above 1 MeV. Thus, as the mission will open a new window, it is difficult to assess what will be discovered. Anyway, we could expect to make detailed studies of jet non-thermal components observed from AGN, stellar black holes (BHs) and GRBs. We might also expect a better description of particle acceleration processes in, for example, pulsars and supernova remnants.

The polarimetry performance of e-ASTROGAM in the Compton range was evaluated with the software package MEGAlib\cite{zog06}, which was originally developed for analysis of simulation and calibration data related to the Compton scattering and pair-creation telescope MEGA\cite{kan05}. The satellite mass model includes passive material in the detectors and their surroundings, true detection thresholds, energy and position measurement accuracy, as well as a roughly accurate spacecraft bus mass and position (see Fig. 17 in Ref.~\citenum{eas} for an illustration of the e-ASTROGAM mass model). To correct the simulated polarization diagrams from possible systematic effects of instrumental origin, for each studied object, we also simulated the azimuthal angular response of the instrument to an unpolarized source with the same spectral distribution and position in the field of view as the polarized source. In practice, the systematic uncertainty in the polarization measurement will be estimated from  the results of an extensive on-ground calibration campaign of the e-ASTROGAM instrument using both polarized and unpolarized monochromatic beams of gamma-rays. 

The dominant sources of background for the e-ASTROGAM telescope in the MeV domain are the cosmic diffuse gamma-ray background, the atmospheric gamma-ray emission, the reactions induced by albedo neutrons, and the background produced by the radioactivity of the satellite materials activated by fast protons and alpha particles. All these components were  modeled in detail using the MEGAlib environment tools. Fluxes of abundant cosmic-ray particles in LEO (protons, alpha particles, electrons, and positrons) were modeled from the works of Mizuno et al. (2004)\cite{miz04} and the cosmic-ray induced atmospheric neutron environment was simulated from the paper by Kole et al. (2015)\cite{kol15}. We also used Ref.~\citenum{miz04}, together with Refs.~\citenum{saz07,chu08,tur10}, to model the cosmic and atmospheric gamma-ray backgrounds for e-ASTROGAM's orbit.

\subsection{Active galactic nuclei}

Many, if not all, galaxies contain a massive central BH, with masses ranging from a million to several billion times that of the Sun,  but not all galaxies containing BHs are currently active.  Specifically, we call a galactic nucleus ``active'' if it contains a bright and compact source of optical radiation (and is often, but not always, detected in other radiation bands). AGN revealing jet-like outflows in the optical band (about 10\% of all AGN) are always strong emitters in the radio band. Blazars are sources where the jet points at (or close to) our line of sight, and the jet radiation being Doppler-boosted can dominate the emission from the accretion disk. Sources possessing a jet which points away from our line of sight often show gamma-ray emission, but at a level significantly weaker than blazars. For sources devoid of jets, the nuclear emission is dominated by thermal or quasi-thermal radiation from the BH accretion disk. The lack of a jet manifests itself by weak or absent emission in the radio and in gamma-rays, and these are Seyfert galaxies, or ordinary ``radio-quiet'' quasars. For a recent overview of taxonomy of gamma-ray emitting AGN, see Ref.~\citenum{mad16}.

The origin and the level of polarization in the hard X-ray/soft gamma-ray band for the radio-quiet sources (presumably associated with the accretion disk) and blazars (presumably associated with the jet) are quite different and are discussed separately.

\subsubsection{Radio-quiet quasars and Seyfert galaxies}

While these sources are copious emitters of soft to mid-range X-rays, they are not strong emitters of gamma-rays in the multi-MeV to GeV range\cite{2012ApJ...747..104A}. The only radio quiet AGN detected by {\it Fermi}/LAT are Seyfert galaxies that happen to be also characterized by strong starburst activity, which in turn correlates with the star-formation rate\cite{2012ApJ...755..164A}, so the accretion disk is not a likely source of GeV gamma-rays.  On the other hand, Seyfert galaxies are often strong emitters of soft gamma-rays, and a number of them were detected by the {\it GCRO}/Oriented Scintillation Spectrometer Experiment\cite{joh97} (OSSE) and subsequently by {\it INTEGRAL}\cite{lub16}. Their spectra extend to several hundred keV, and as such, they are likely to be detected at the lower energy range of e-ASTROGAM:  good bright examples are the Seyfert 1 galaxies NGC\,4151 and IC\,4329a, and the Seyfert 2 galaxy NGC\,4945, all of them detected at $\sim 200$ keV (but with steep spectra, falling with energy).  In all cases, the X-ray and soft gamma-ray emission is often seen to be variable (on time scales of roughly days, depending on the  source), and such variability constrains the size of the X-ray emission region.

Our current best model for the formation of the X-ray spectra in these sources is Compton upscattering of the soft (UV-range) accretion disk photons -- generated via quasi-thermal dissipation in the disk -- to X-ray energies by a trans-relativistic plasma located as a corona above the accretion disk, or possibly some axially-symmetric proto-jet.  The details of the geometry of the Compton-scattering medium are not known, but one would expect that any such sources -- when viewed away from the axis of symmetry -- would show polarization (because of the angular distribution of Compton scattering), and the angle of polarization would provide additional, important hints regarding the geometry of the Comptonizing medium.  This is in addition to the crucial information about the temperature of the Comptonizing plasma, which can be gleaned from shape of the emission spectrum in the X-ray to gamma-ray range.  

\subsubsection{Jet-dominated active galaxies (blazars)}

Broad-band emission from blazars is quite different to that from radio-quiet, accretion-dominated sources. Here, the emission extends to much higher energies:  blazars are generally detected throughout the {\it Fermi} band (with clear detections in the 0.05 - 50 GeV range).  The radio emission is extremely compact (with structures as small as milli-arcsecond, corresponding to sub-parsec scales), and is strongly polarized.  The optical emission is highly variable, and often also strongly polarized.  The optical spectra often (but not always) show emission lines, most clearly discernible when the sources are optically faint. Without exception, blazars are sources of X-ray emission. Measurement of polarization in the X-ray to gamma-ray range provides crucial insight not only to the geometry of the emitting region, but also allows to discriminate among various processes proposed as emission mechanisms.

\begin{figure}
\begin{center}
\begin{tabular}{c}
\begin{minipage}{0.435\linewidth}
\includegraphics[scale=0.35]{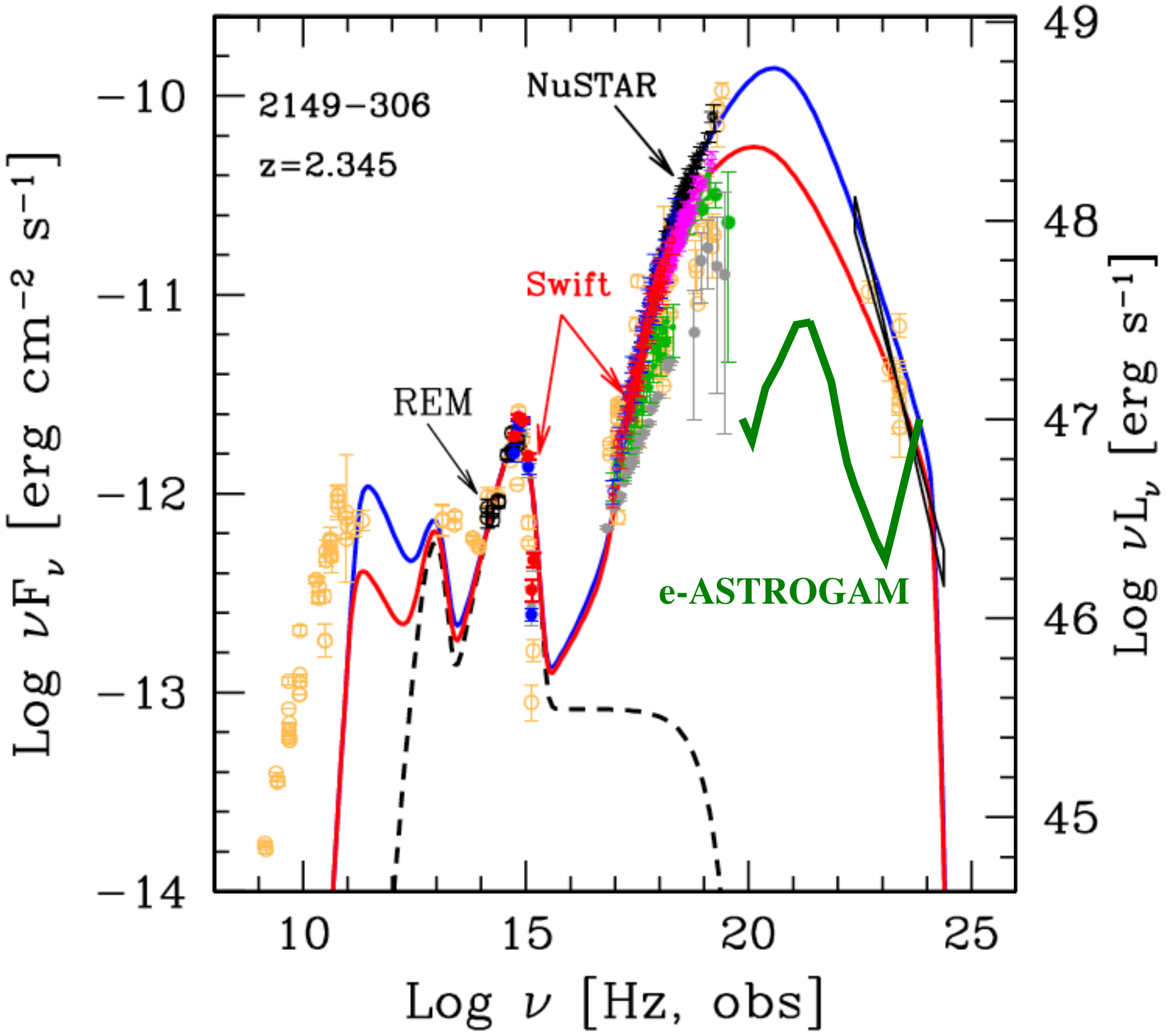}
\end{minipage}
\begin{minipage}{0.565\linewidth}
\includegraphics[scale=0.35]{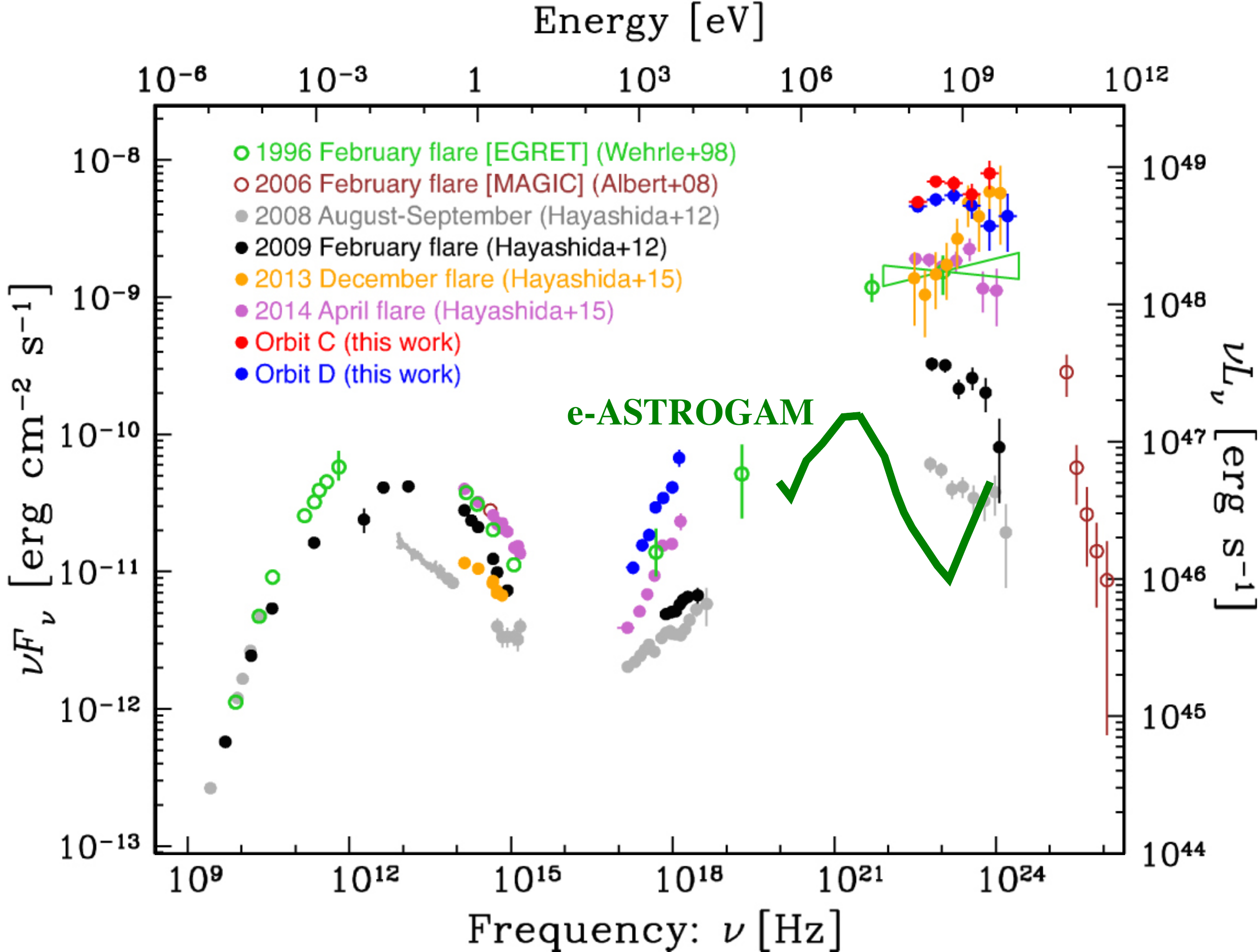}
\end{minipage}
\\
(a) \hspace{8.cm} (b)
\end{tabular}
\end{center}
\caption 
{ \label{fig:agn}
Spectral energy distributions of two FSRQ blazars together with the e-ASTROGAM sensitivity. (a) PKS 2149-306, which is at a redshift of 2.345, and the model of Ref.~\citenum{tag15}. (b) A collection of different spectral states of 3C 279 showing the dramatic gamma-ray flaring activity of this blazar\cite{2016ApJ...824L..20A}. The e-ASTROGAM sensitivity is for a source exposure of 1 yr in panel (a) and 1 day in panel (b).} 
\end{figure} 

For most blazars, their spectra, when plotted in the $\nu F_\nu$ representation, show two prominent broad humps, one peaking (depending on the source) at mm-to-soft X-ray band, and the other peaking in the gamma-ray band (Figure~\ref{fig:agn}). We note that the mid-range gamma-ray emission is poorly studied, owing to the limited sensitivity of instruments flown so far. High levels of polarization measured for the low-energy hump (see, e.g., Refs.~\citenum{mea90,hay12}), often exceeding 20\% -- together with frequently observed rapid polarization variability -- argues for the emission process to be via the synchrotron mechanism.  The high-energy peak is thought to originate via the inverse Compton process, presumably by the same electrons that produced the synchrotron hump.  All this takes place in a volume that can be estimated from variability time scales.  Broad-band spectra of blazars are successfully modeled by such synchrotron + inverse Compton models.  However, while generically such a model appears to be applicable for all blazars, there are differences depending on the luminosity of the blazar.

In the luminous objects -- often characterized by prominent optical and UV emission lines, and known as Flat Spectrum Radio Quasars, or FSRQs -- the source of ``seed'' photons for Compton scattering is likely to be from the direct or reprocessed emission from the accretion disk, and most likely is provided by the optical/UV emission line-producing clouds (Ref.~\citenum{sik94}).  In addition, the diffuse infrared radiation provided by the torus is also likely to be a significant source of ``seed'' photons\cite{sik09}.  Such radiation is not expected to be polarized.  However, especially at the lower end of the high-energy peak (meaning in the X-ray band), the source of ``seed'' photons might be the synchrotron emission in the jet, as is the case for low luminosity objects.

In low-luminosity sources, the emission from the accretion disk and/or emission line clouds is weak (often undetectable).  Those are the lineless BL Lac objects and the dominant source of the ``seed'' photons for inverse Compton scattering is most likely the synchrotron radiation produced in the jet.  

A good overview of the polarization signature expected in the hard X-ray/soft gamma-ray bands is provided in Ref.~\citenum{kra12}.  The level of polarization depends obviously on the emission process responsible for radiation in the band under consideration.  For FSRQ-type blazars, the e-ASTROGAM 0.15 - 3000 MeV range corresponds, in all current scenarios, to the inverse Compton peak (see Figure~\ref{fig:agn}a).  If the ``seed'' photons are polarized, it is expected that the inverse Compton emission of photons upscattered via Comptonization should also show some level of polarization, but probably less than the synchrotron photons emitting in the optical/UV band. Specifically, authors of Ref.~\citenum{kra12} suggest that for the case of an FSRQ-type object radiating soft gamma-rays via the synchrotron self-Compton process, the polarization level should be as much as 30 to 40\% (but, realistically, somewhat lower, because of tangled nature of magnetic field). If those ``seeds'' are not polarized, and the source and the scatter are extended, there should be no discernible polarization in the up-scattered photons.  Here, the measurement of polarization would be the ``smoking gun'' for the origin of the ``seeds'' for Comptonization.  Polarization of hard X-rays/soft gamma-rays would implicate the synchrotron process internal to the jet.  If no polarization is detected -- this would support the radiation external to the jet. 

An intriguing possibility is that the rapidly variable component of gamma-ray emission during rapid and luminous outbursts of FSRQ blazars (such as 3C 279; see Figure~\ref{fig:agn}b) might be synchrotron emission from the most energetic electrons (Refs.~\citenum{nal12,2016ApJ...824L..20A}). In that case, the polarization of gamma-ray emission during such outbursts should be very strong and readily measurable by e-ASTROGAM. Its level should be comparable to the polarization measured in the optical band, also produced by the synchrotron process, which during blazar flares can be as large as 40\%.  It is important to note that the optical polarization level can be variable, and often is associated with rotation of the polarization angle, which, in turn commonly is associated with gamma-ray flares (see, e.g., Refs.~\citenum{mar08,abdo10,hay12};  for a recent monitoring of polarization level in blazars, see Refs.~\citenum{bli16,ito16}).  An additional diagnostic can be obtained by correlating the time series of the angle of optical and gamma-ray polarization vector, as suggested in Ref.~\citenum{kra12}. 

For BL Lac objects, hard X-rays are probably still dominated by the ``tail'' of the synchrotron process, so the polarization should be strong.  Since there are very few measurements of BL Lac blazars in the spectral region where the e-ASTROGAM is sensitive, it is not clear where the ``onset'' of the high energy, inverse-Compton hump is located.  If our current model for the  broad-band spectra of BL Lacs is correct, one should see the broad-band spectra falling sharply (in $\nu F_\nu$) up to several hundred keV, and rising again at even higher energies. We should see strong polarization associated with the falling spectrum (photon index $> 2$), weakening (but still discernible) at the point where the spectrum breaks to photon index $< 2$. A more extensive and quantitative discussion is in papers by Chakraborty et al.\cite{cha15} and Zhang et al.\cite{zha13,zha16}, but qualitatively, the general picture outlined above is in agreement with those papers. In addition, Zhang et al.\cite{zha13,zha16} have pointed out that gamma-ray polarization can provide definitive insight into the presence of hadrons in blazar jets, because hadronic models predict a much higher degree of polarization than leptonic models. 

In summary, the sensitive spectroscopy and measurement of polarization in the e-ASTROGAM spectral range is crucial towards the understanding of emission models in all classes of AGN.

\subsection{Gamma-ray bursts}

Gamma-ray Bursts (GRBs) are the most luminous electromagnetic transients in the universe, with spectra that peak in the 0.1--1\,MeV range. Their huge luminosities ($\sim 10^{51}$ erg in a narrow collimated jet) are emitted by the most relativistic jets known ($\Gamma > 100$) and are observed over a very short timescale (tens of seconds) at redshifts extending up to at least z\,$\sim$\,9. The `short' GRBs, with durations $\lsim 2$\,s are believed to originate from compact binary mergers (neutron star -- neutron star, neutron star -- BH), while those with durations $\gsim 2$\,s (`long' GRBs ) originate from the collapse of massive stars.  GRBs are therefore powerful probes of the early universe, of fundamental physics, and of matter and radiation under the most extreme conditions\cite{geh13,ber14}.  

However, the nature of the central engines producing the relativistic jets in GRBs, the physical properties of the jets themselves, the energy dissipation sites and the radiation mechanisms, all remain poorly understood. In the BH scenarios, the jet may be launched by the Blandford-Znajek mechanism, provided that sufficiently strong magnetic fields thread the BH horizon. If the magnetic field does play an important role as an energy mediator in jet production, then a highly magnetized outflow close to the jet-launching site is expected and such a jet would be magnetically dominated. In the case of synchrotron emission with ordered magnetic fields, the observed GRB polarization is expected to vary little with viewing angle. However, in the case of synchrotron emission in random magnetic fields and for Compton-drag models,  the observed polarization should be significant only near the edge of the jet. These families of models can be distinguished by collecting a sufficiently large sample of GRBs with accurately measured polarization fractions\cite{tom09}. 

In GRBs the relativistic jets are optically thick in the launch region\cite{lun14}. As the jets expand and become transparent, they release internally trapped photons as photospheric emission that has a distinctive polarization signature, which evolves during the GRB. Polarisation measurements as a function of energy and time can therefore discriminate between synchrotron and photospheric emission mechanisms in the prompt GRB phase. 

SPI and IBIS (Imager on Board the INTEGRAL Satellite) on \textit{INTEGRAL} have demonstrated the capability to detect the signature of polarized emission in the medium energy gamma-ray range from $\gamma$-ray sources, including GRBs. For example, GRB\,041219a was an intense burst localized by \textit{INTEGRAL}, whose degree of linear polarization in the brightest pulse of duration 66\,s was found by SPI to be
$63^{+31}_{-30}$\% at an angle of $70^{+14}_{-11}$~degrees in the 
100--350~keV  energy range \cite{mcglynn07}. The same interval analysed using IBIS Compton mode data yields an upper limit on the polarisation fraction of  $\leq 4\%$. The IBIS data for the intense first pulse of this exceptionally bright GRB was heavily affected by telemetry saturation, which may explain the disparity between the results from the two instruments. The polarisation results from SPI and IBIS for the brightest 12\,s interval of the first pulse in this GRB are consistent with each other \cite{gotz09}. 
The relatively low significance of these detections ($\sim 3\sigma$), combined with the lack of on-ground calibration of polarimetric response to quantify systematic effects, limit the  scope of what can be achieved with instruments that were not expressly designed for polarization measurements.   

The Gamma-Ray Burst Polarimeter (GAP) aboard the Small Solar Power Sail Demonstrator, IKAROS, launched in 2010, was the first instrument designed and calibrated for polarimetric GRB observations between 50--300\,keV\cite{yon11a}.  GAP notably detected GRB\,100826A, with an average polarization degree of 27\% $\pm$ 11\%, 2.9\,$\sigma$ confidence level\cite{yon11b}. The polarisation angle was observed to vary by $\sim$\,90$^{\circ}$ during the prompt emission, which is difficult to explain for an axi-symmetric jet in the above scenarios. The polarisation swing during the burst can be interpreted as evidence for patchy emission, patchy magnetic field, or as photospheric emission from a variable jet. A further 2 GRBs were observed by GAP to have polarization degrees above 70\%, but with no change in polarization angle\cite{yon12}, consistent with magnetic field structures in the emission region being globally ordered fields advected from the central engine.

Several other soft $\gamma$--ray polarimetric instruments have been proposed, and are in various stages of implementation. For example, Polarimeters for Energetic Transients (POET) has been proposed to NASA as a SMEX mission with a science payload composed of two polarimetry instruments to make polarization and spectral measurements from about 2\,keV up to about 500\,keV\cite{mcc14}. POET's High Energy Polarimeter (HEP) instrument design derives from the balloon-borne polarimeter GRAPE  (Gamma-RAy Polarimeter Experiment) that was successfully flown in 2011 and placed constraints on the polarization from the Crab Nebula and two M-class solar flares\cite{con10}. POLAR is a dedicated GRB polarimeter that has been proposed as a payload for the Chinese Space Laboratory. It is capable of detecting $\sim 10$\,GRBs/yr with minimum detectable polarization (MDP) $\leq$\,10\% \cite{xio09}. The Astrosat mission, launched in September 2015, has a cadmium zinc telluride (CZT) imager\cite{vad15} with the capability to make time resolved polarisation measurements for 5--6 GRBs per year\cite{rao16}. 

\begin{figure}
\begin{center}
\begin{tabular}{c}
\begin{minipage}{0.5\linewidth}
\includegraphics[scale=0.4]{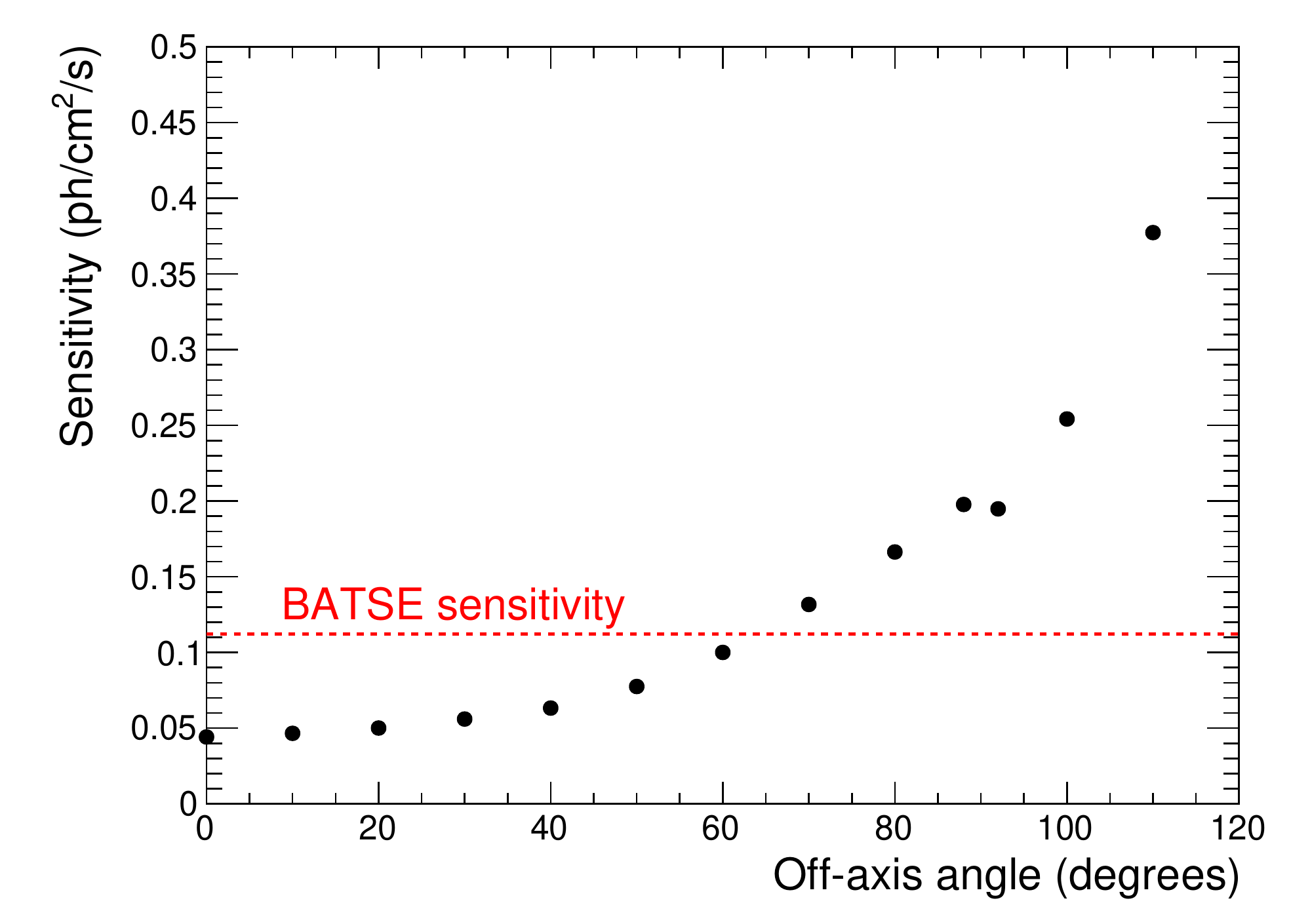}
\end{minipage}
\begin{minipage}{0.5\linewidth}
\includegraphics[scale=0.4]{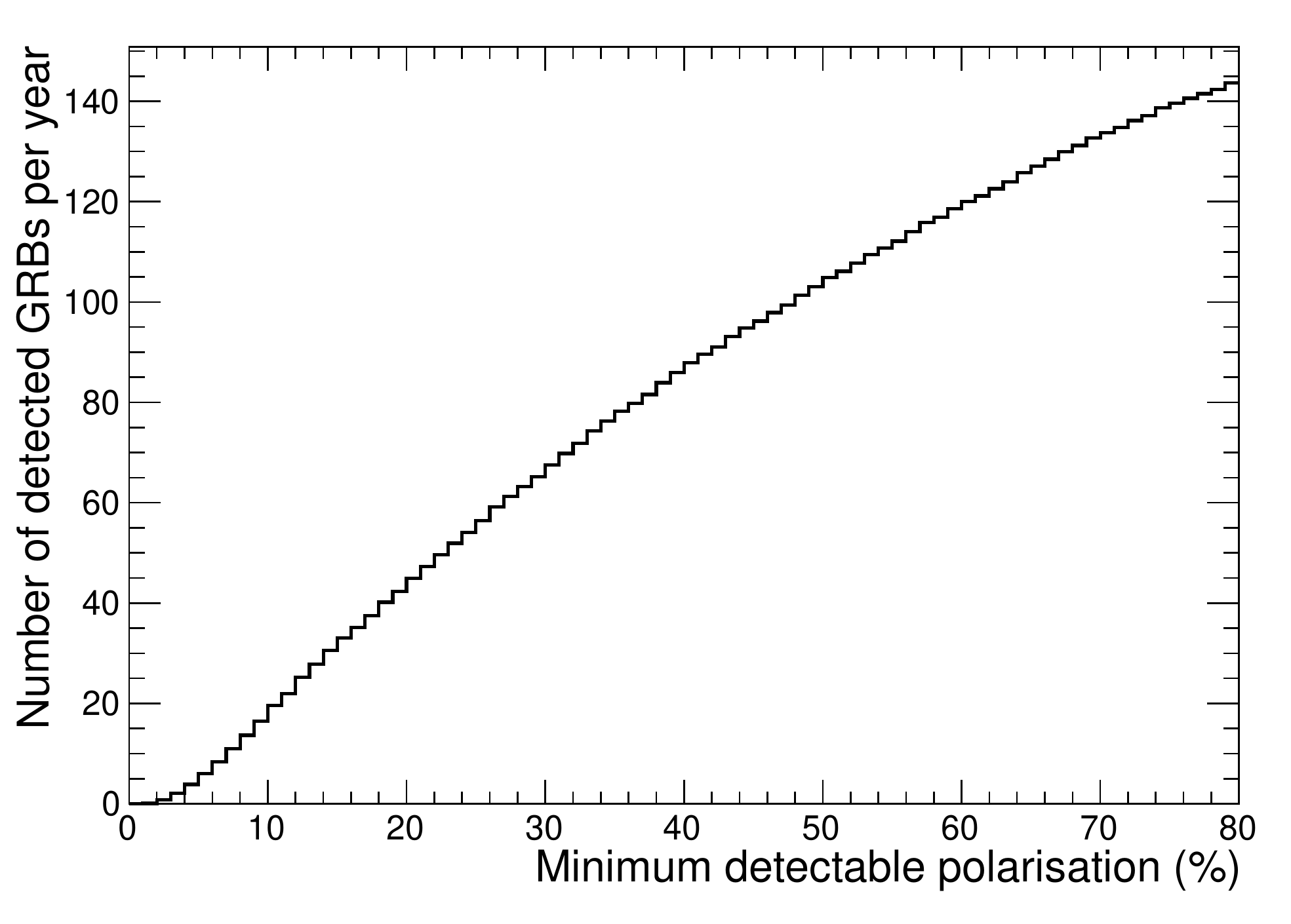}
\end{minipage}
\\
(a) \hspace{8.2cm} (b)
\end{tabular}
\end{center}
\caption 
{ \label{fig:GRB}
(a) GRB flux sensitivity of e-ASTROGAM in the 0.2--2\,MeV range on a timescale of 1\,s. No data point is shown for an off-axis angle of $90^\circ$ because the trajectories of incident gamma rays in that case are parallel to the DSSD planes, resulting in a loss of effective area and sensitivity. The dashed line shows the sensitivity of the BATSE instrument in the same energy range, which was calculated from its sensitivity of 0.29 photon cm$^{-2}$ s$^{-1}$ in the 50--300\,keV range assuming a typical GRB spectrum described in the text. (b) Cumulative number of GRBs to be detected by e-ASTROGAM as a function of the minimum detectable polarization MDP$_{99}$ in the 150--300\,keV range.} 
\end{figure} 

 The left panel of Figure~\ref{fig:GRB} shows the expected sensitivity of e-ASTROGAM for the detection of GRBs in the 0.2--2\,MeV energy range, as a function of the off-axis angle $\theta_{\rm off-axis}$. Here, the GRB emission spectrum has been approximated by a typical Band function \cite{ban93} with $\alpha=-1.1$, $\beta=-2.3$, and $E_{\rm peak}=0.3$~MeV. The e-ASTROGAM sensitivity is expected to be better than that of {\it CGRO}/BATSE for $\theta_{\rm off-axis} \lsim 65^\circ$. Overall, the total number of GRBs detected by e-ASTROGAM is expected to be $\sim$600 in 3 years of nominal mission lifetime. 

The strength of e-ASTROGAM regarding GRB polarimetry will be in its reliable polarization measurements for a large sample.
The right panel of Figure~\ref{fig:GRB} shows the number of GRBs detectable by e-ASTROGAM as a function of the minimum detectable polarization at the 99\% confidence level (MDP$_{99}$) in the 150--300\,keV band. The response of e-ASTROGAM to linearly polarized GRBs has been simulated at several off-axis angles in the range $[0^\circ;90^\circ]$. For each value of $\theta_{\rm off-axis}$, the distribution of reconstructed azimuth scatter angles has been corrected for the asymmetry of the detector acceptance using a similar distribution obtained for an unpolarized source. The minimum detectable polarization has been calculated as
\begin{equation}
\mathrm{MDP}_{99}=\frac{4.29}{\mu_{100}S}\sqrt{S+B},
\end{equation}
where $\mu_{100}$ is the modulation amplitude for a 100\% polarized source, $S$ and $B$ are the number of source and background counts, respectively~\cite{weisskopf}. The number of GRBs with polarization measurable with e-ASTROGAM has then been estimated using the GRB fluences and durations from the Fourth BATSE GRB Catalog \cite{pac99}. The minimum detectable polarization has been calculated for each GRB in the catalogue using five off-axis angles ($26^\circ$, $45^\circ$, $60^\circ$, $72^\circ$ and $84^\circ$) with an event weight of 0.1 to approximate the uniform distribution of GRBs over the full sky (GRBs with angles over $90^\circ$ have been discarded due to low sensitivity; the telescope is assumed to be zenith-pointing). From Figure~\ref{fig:GRB} it can be seen that e-ASTROGAM should detect a polarization fraction of 20\% in about 40 GRBs per year, and a polarization fraction of 10\% in $\sim$16 GRBs per year. 

\begin{figure}
\begin{center}
\begin{tabular}{c}
\begin{minipage}{0.5\linewidth}
\includegraphics[scale=0.4]{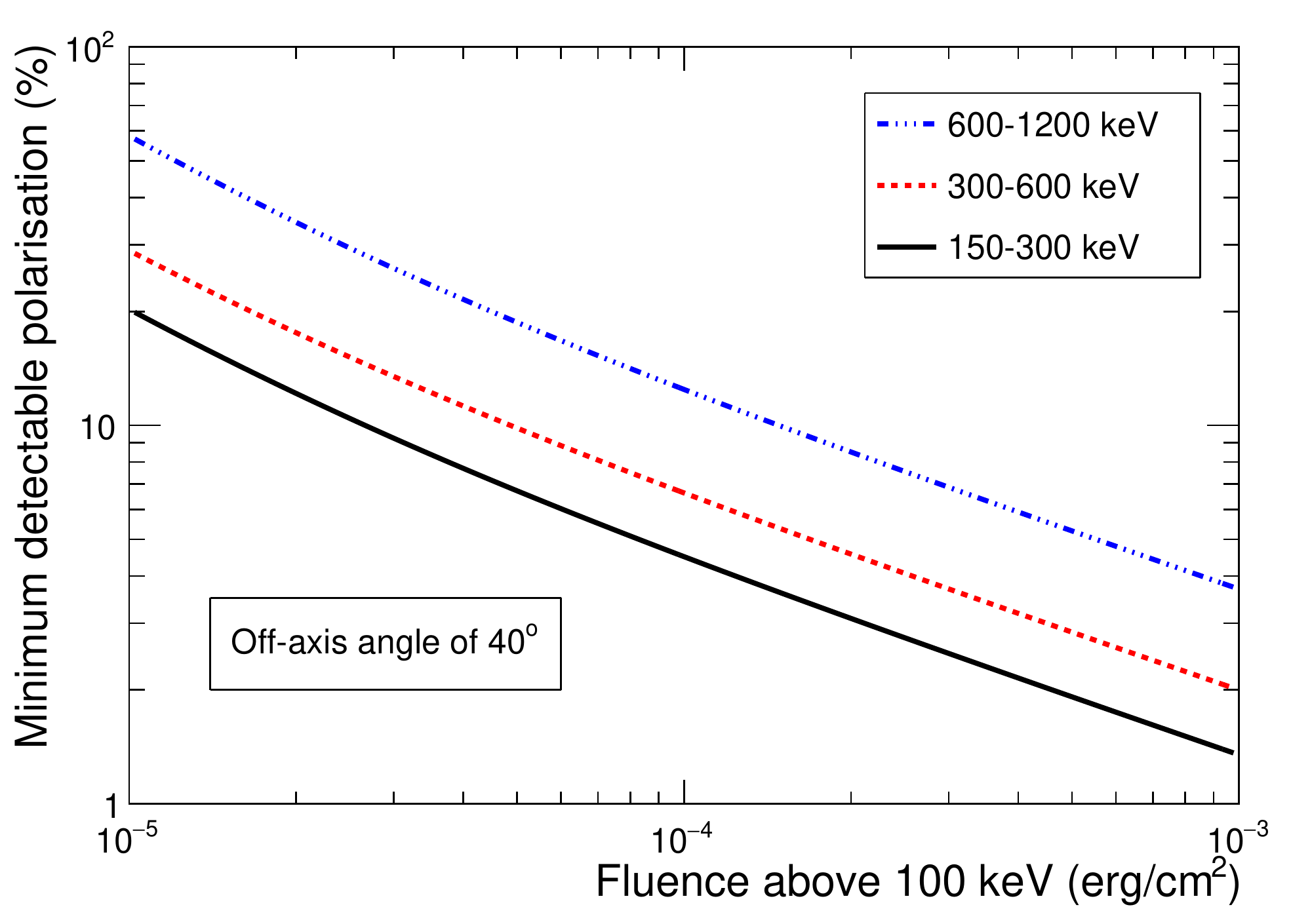}
\end{minipage}
\begin{minipage}{0.5\linewidth}
\includegraphics[scale=0.4]{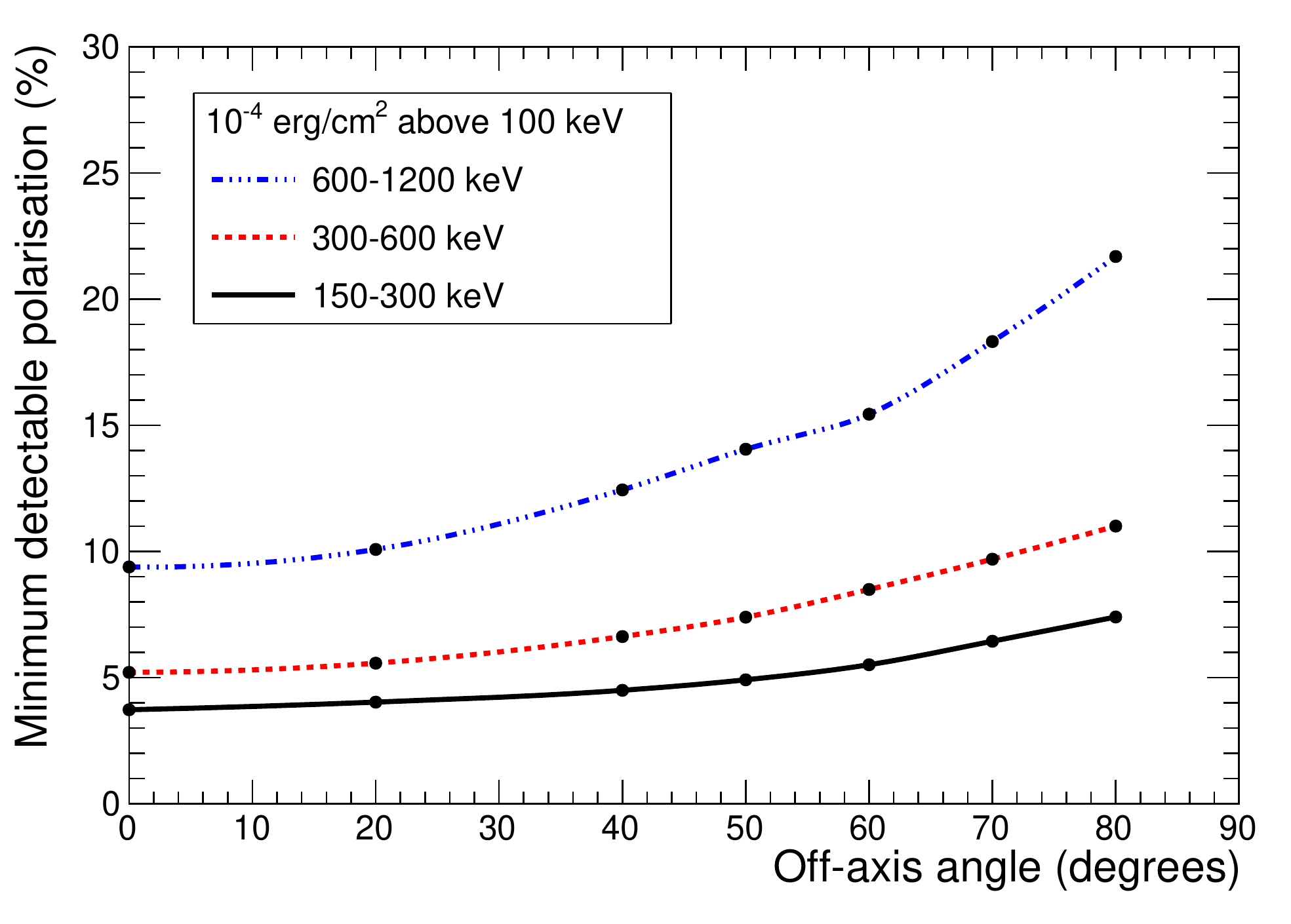}
\end{minipage}
\\
(a) \hspace{8.2cm} (b)
\end{tabular}
\end{center}
\caption 
{ \label{fig:GRB_MDP}
Minimum polarization fraction MDP$_{99}$ detectable by e-ASTROGAM in three energy bands as a function of the GRB fluence (a) and as function of the off-axis angle (b). The GRB duration is assumed to be 50~s and the spectrum is a Band function with $\alpha=-1.1$, $\beta=-2.3$, and $E_{\rm peak}=0.3$\,MeV.} 
\end{figure} 

For bright GRBs (with fluence above 100 keV $\gsim 10^{-4}$~erg cm$^{-2}$), e-ASTROGAM will be able to measure polarization for several energy bands between 150 and 1200~keV (Figure~\ref{fig:GRB_MDP}). This polarization information, combined with spectroscopy over a wide energy band, will provide unambiguous answers regarding the origin of the GRBs' highly relativistic jets and the mechanisms of energy dissipation and high-energy photon emission in these extreme astrophysical phenomena.

\subsection{The Crab pulsar and nebula}

The Crab nebula is the remnant of a star that exploded in 1054. The leftover neutron star, the Crab pulsar, has a rotation period of 33 milliseconds and spreads high-energy particles, up to 10$^{15}$~eV, into the surrounding medium. Thanks to the energetic population of electrons in the pulsar wind, the compact nebula emits at all wavelengths, in particular at MeV energies by synchrotron radiation and at higher energies via inverse Compton scattering. The bright synchrotron radiation is particularly well suited for detailed studies of the inner region of the Crab pulsar/nebula system. For decades the flux from the whole nebula was expected to be constant and consequently this ``standard candle'' was observed by numerous space experiments for high-energy calibration purposes. This situation has changed in 2007 with the detection by the {\it AGILE} and {\it Fermi} missions of strong $\gamma$-ray flares\cite{fermi11,tav11,striani13}. Nevertheless, the precise localization of the emission inside the nebula and the physical origin of this sudden bright activity are not yet well understood, requiring more studies. 

Polarized emission from the nebula and pulsar has been measured from radio to X-rays up to a few keV\cite{weiss78}. Recently, polarization in hard X-rays was detected with the two main instruments on board the {\it INTEGRAL} satellite, IBIS and SPI\cite{forot08,dean08}. These polarimetric measurements give access to information related to the wind geometry and the magnetic field configuration in the pulsar environment, parameters that make it possible to constrain where and how efficiently particle acceleration takes place. In the MeV range, both the pulsar and nebular components of the system contribute significantly to the total emission. Thanks to the excellent sensitivity and timing resolution of the e-ASTROGAM telescope, phase-resolved polarimetric analysis will be possible, allowing a detailed study of the compact inner nebula surrounding the Crab pulsar (the so-called pulsar wind nebula).

\begin{figure}
\begin{center}
\begin{tabular}{c}
\begin{minipage}{0.5\linewidth}
\includegraphics[scale=0.32]{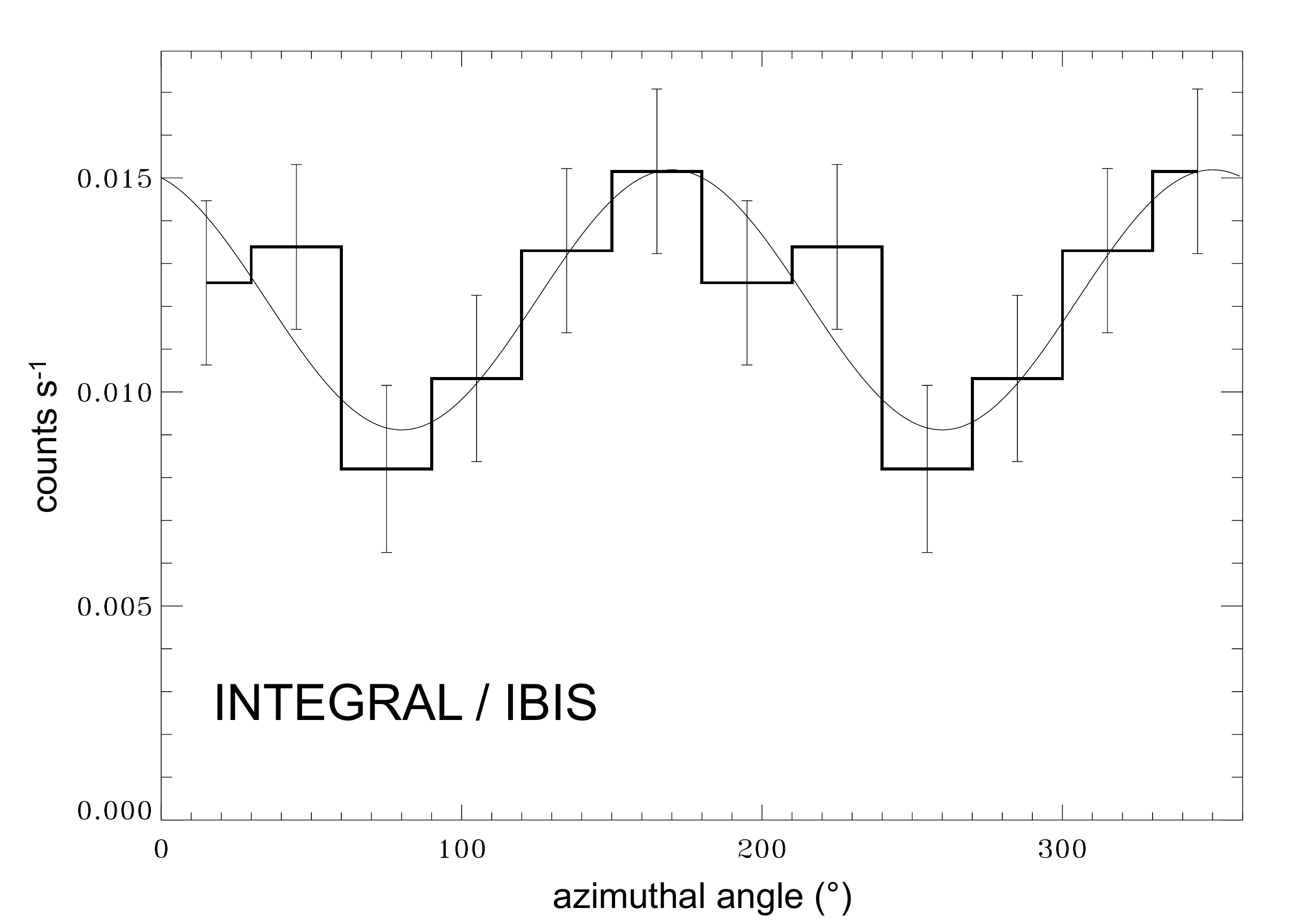}
\end{minipage}
\begin{minipage}{0.5\linewidth}
\includegraphics[scale=0.31]{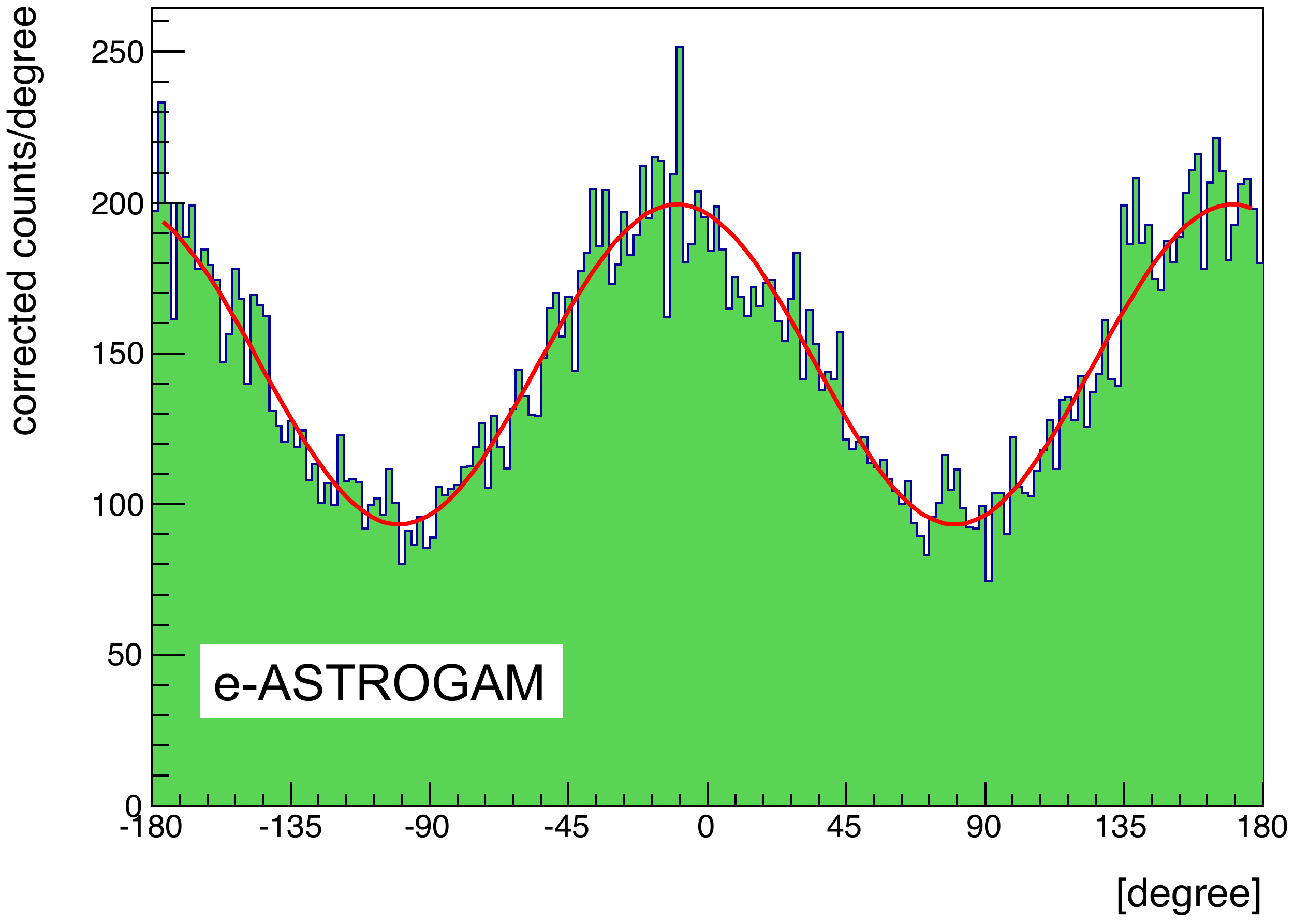}
\end{minipage}
\\
(a) \hspace{8.2cm} (b)
\end{tabular}
\end{center}
\caption 
{ \label{fig:Crab}
(a) Phase-averaged polarization diagram of the Crab pulsar and nebula obtained with {\it INTEGRAL}/IBIS in the 300 -- 450 keV energy band by summing 1.831~Ms of data taken between 2012 and 2014 (adapted from Ref.~\citenum{mor16}). The measured polarization fraction and angle are ${\rm PF}=(98\pm37)$\% and ${\rm PA}=80^\circ \pm 12^\circ$, respectively. (b) Simulation of the e-ASTROGAM polarization signal in the same energy band (300 -- 450 keV) for an effective duration of observation of the Crab pulsar and nebula 100 times shorter than that accumulated by {\it INTEGRAL}/IBIS (i.e. $T_{\rm obs} = 1.831 \times 10^4$~s $ \simeq 5$~h). The reconstructed polarization fraction and angle are ${\rm PF}=(98.0 \pm 2.4)$\% and ${\rm PA}=80.2^\circ \pm 0.7^\circ$.} 
\end{figure} 

Figure~\ref{fig:Crab} shows a comparison of a phase-averaged polarization diagram for the Crab pulsar and nebula obtained with {\it INTEGRAL}/IBIS\cite{mor16} summing 1.831~Ms of data between 2012 and 2014 and a simulated polarigramme for e-ASTROGAM observation of the Crab region with 100 times shorter exposure ($T_{\rm obs} = 1.831 \times 10^4$~s). For the simulations we used the Crab energy spectrum of Ref.~\citenum{jou09} and assumed the central values measured with {\it INTEGRAL}/IBIS\cite{mor16} for the polarization fraction (${\rm PF}=98.0$\%) and angle (${\rm PA}=80.0^\circ$) of the soft gamma-ray emission ($\Delta E = 300$~--~$450$~keV). The improvement of the data quality with e-ASTROGAM is obvious, as evidenced by the large increase in the number of bins made possible by the much higher statistics accumulated with e-ASTROGAM (with 100 times less exposure). 

\begin{figure}
\centering
\includegraphics[width=0.55\textwidth]{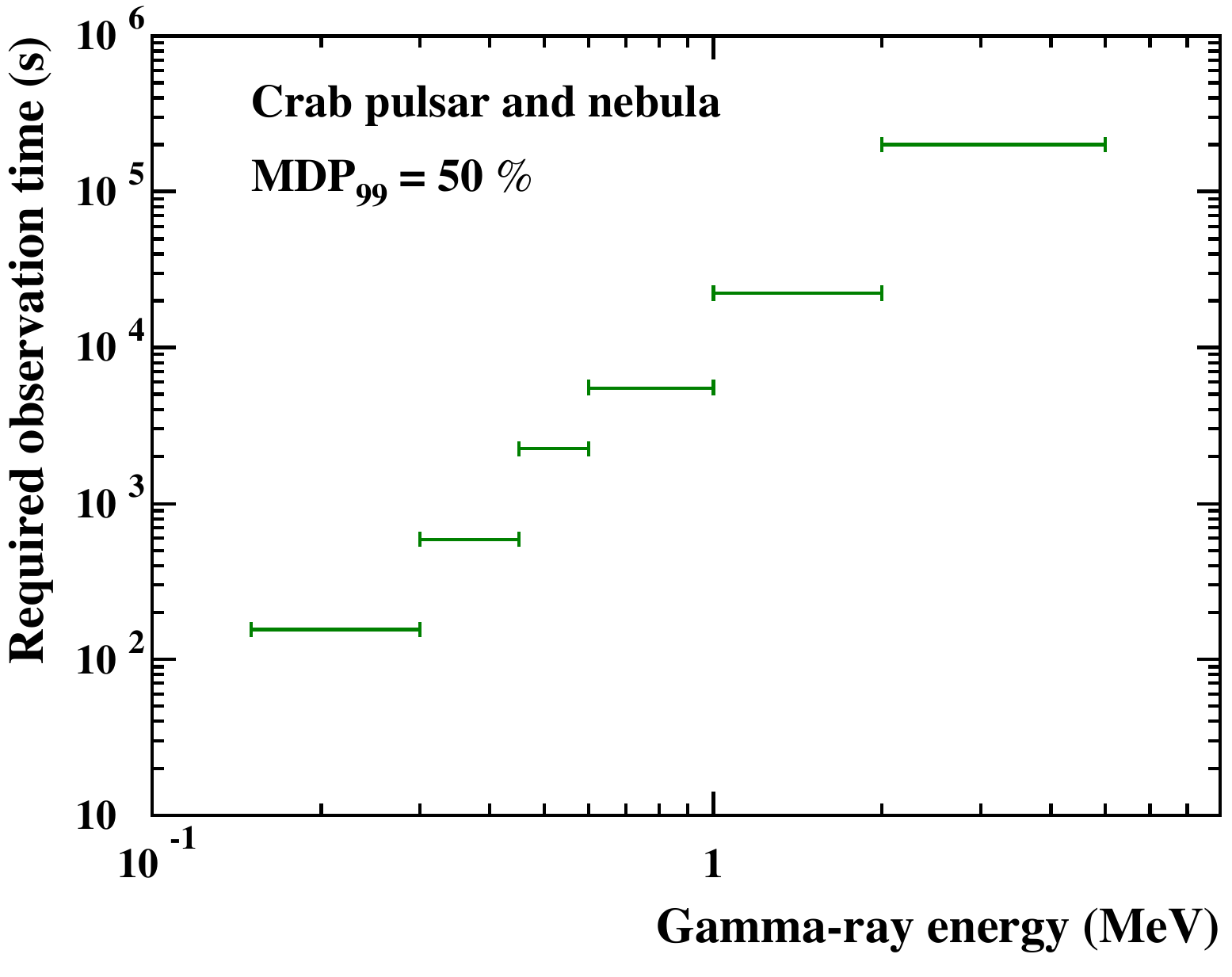}
\caption{Required e-ASTROGAM observation time to detect a polarization fraction of 50\%  at the 99\% confidence level from a Crab-like source in six energy bands, 150 -- 300 keV, 300 -- 450 keV, 450 -- 600 keV, 0.6 -- 1 MeV, 1 -- 2 MeV, and 2 -- 5 MeV, using Compton interactions in the instrument. The polarimetry capability of e-ASTROGAM at higher energies, using pair production events, is not shown, as it requires more detailed simulations.}
\label{fig:Crab_Tobs}
\end{figure}

Figure~\ref{fig:Crab_Tobs} shows the required time of observation with e-ASTROGAM to detect a polarization fraction of 50\% from a Crab-like source in various energy bands. We see that in the $300$~--~$450$~keV band studied by INTEGRAL (Figure~\ref{fig:Crab}) the required observation time with e-ASTROGAM is only of 10~min, which means that a total exposure of 2 Ms would allow a phase-resolved polarimetric study of the Crab pulsar with a time resolution of 10~$\mu$s (i.e. $1/3300$ of the rotation period). Up to 2 MeV, a time-resolved analysis can be carried out with a resolution better than 400~$\mu$s (see Figure~\ref{fig:Crab_Tobs}). Thus, e-ASTROGAM will be able to provide a detailed gamma-ray view of the time-dependent polarization properties of the Crab nebula and pulsar, in particular during the periods of strong gamma-ray flaring activities possibly associated with magnetic reconnection \cite{tav11,fermi11}. 

\begin{figure}
\centering
\includegraphics[width=0.7\textwidth]{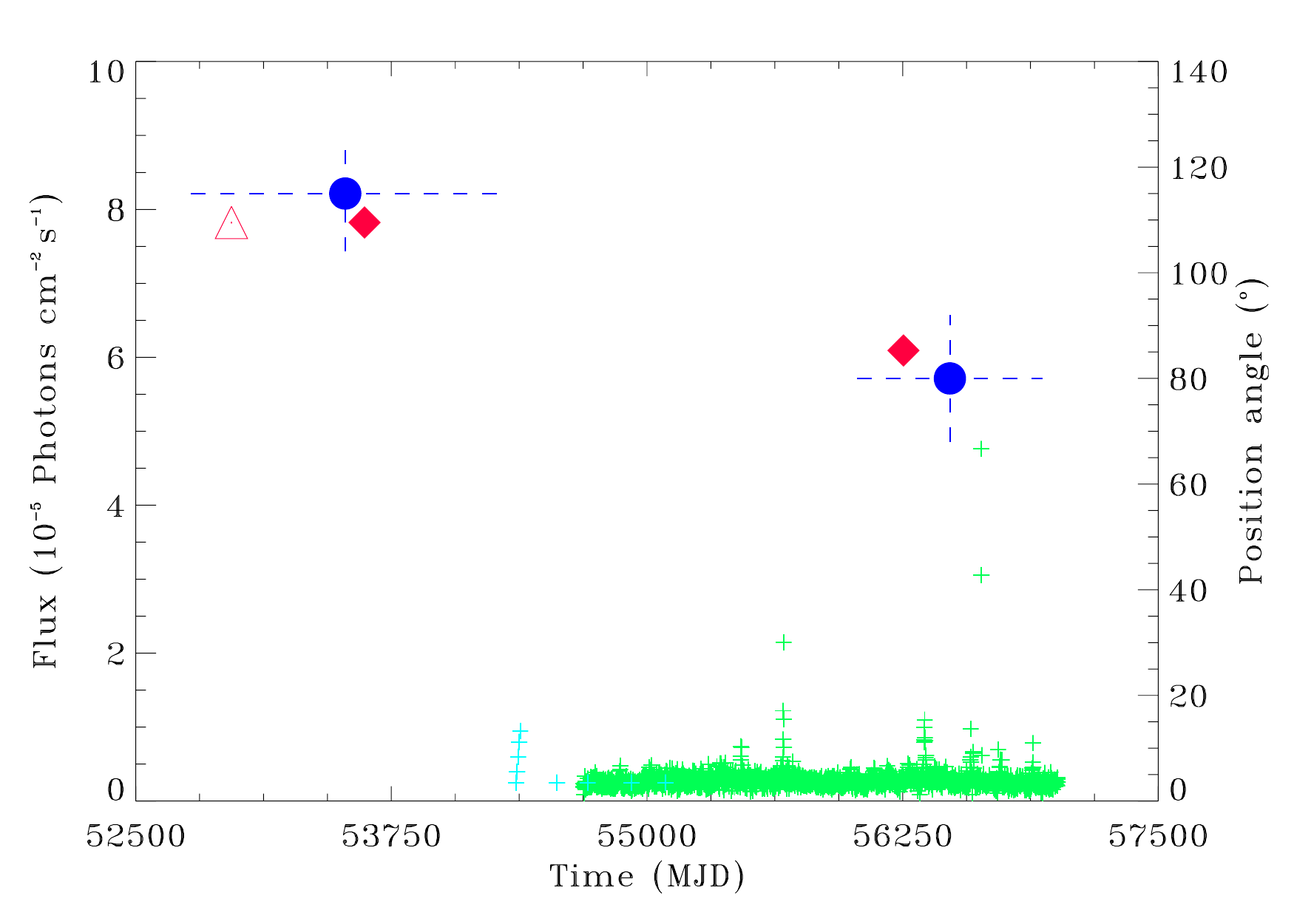}
\caption{Crab light curve in the high-energy band (above 100 MeV) measured by {\it Fermi} and {\it AGILE} (green and cyan color) and polarization measurements obtained between 2003 and 2014 in the optical (red points) and in hard X-rays (300--450 keV) with {\it INTEGRAL}/IBIS (blue filled circles)\cite{mor16}. A similar change in the polarization position angle appears in both bands. e-ASTROGAM will determine the polarization properties of the system in the MeV range on a timescale similar to that of optical observations ($\sim$hours).}
\label{fig:Crab_IBIS-GASP}
\end{figure}

Combined polarimetric observations of the Crab pulsar/nebula system with {\it INTEGRAL} in hard X-rays and in the optical band with the Galway Astronomical Stokes Polarimeter (GASP) optical polarimeter have led to the suggestion that the polarization angle of both the visible and high energy light changed in a similar way over a year-long timescale\cite{mor16}. This result was tentatively attributed to magnetic reconnection events as observed in the solar corona or detected in AGN\cite{abdo10} during bright high-energy flares. The e-ASTROGAM performance will definitively permit simultaneous MeV-optical observations, allowing a detailed study of these variations on a shorter timescale (Figure~\ref{fig:Crab_IBIS-GASP}). 

Polarimetric studies with e-ASTROGAM will help to better understand particle acceleration near compact objects, their injection and interaction in their close environment and the geometry and structure of the magnetic field in these systems. In coordination with highly-promising future facilities in the optical and radio bands (E-ELT and SKA), the Crab pulsar/nebula system is the ideal target for such studies. 

Beyond this prime objective, the e-ASTROGAM mission will make it possible to study the polarimetric properties of pulsars and associated nebulae with different characteristics (e.g. energetics, age and magnetic field strength), including, for example, the pulsar wind nebula MSH\,15-52 (PSR B1509-58) in our Galaxy and the pulsar within the young supernova remnant SNR\,0540-69.3 in the Large Magellanic Cloud. The pulsar PSR B1509-58 is about three times weaker than the Crab above 1~MeV, as found with {\it CGRO}/COMPTEL\cite{kui99}, so it will need about ten times more exposure. PSR\,0540-69.3 was not detected with {\it INTEGRAL} above $200$~keV\cite{slo07}, but it is however interesting, because it is the brightest extragalactic pulsar we know and it is similar to the Crab.

\subsection{Microquasars}

Black Hole Binaries (BHB) transit through different ``spectral states'' during their outbursts, the two canonical ones are the soft state (SS) and the ``low'' hard state (LHS). In the SS the emission is dominated by a bright and warm ($\sim 1$ keV) accretion disk, the level of variability is weak, and the power density spectrum (PDS) is power-law like. No radio emission is detected in this state, and this is interpreted as an evidence for an absence of jets. In the LHS, on the other hand, the disk is much colder ($\sim 0.5$ keV) and is thought to be truncated at a large distance from the accretor. The X-ray spectrum also shows a strong power-law like component extending up to hundred of keV, sometimes showing a roll-over at typically 50~--~200 keV. The level of rapid variability is much higher and the PDS shows a band-limited noise component, with sometimes quasi-periodic oscillations with frequencies in the range 0.1~--~10 Hz. During this state a ``compact-jet'' is detected mainly through its emission in the radio to infrared domain.

The BHB Cygnus X-1 (Cyg X-1) was discovered in 1964 \cite{Bow65} and is the brightest binary in the sky, detectable in the MeV range. It is the first Galactic source known to host a BH\cite{Bol75}, and the most recent estimates led to a BH mass of $15 \pm 1$ solar masses\cite{oro11,zio14}. Cyg X-1 belongs to the family of microquasars owing to the detection of compact relativistic jets in the LHS \cite{Sti01}, and it is one of the (very few) microquasars known to have a hard tail extending to (and beyond) the MeV range (see Ref.~\citenum{rol17} and references therein). This has been recently confirmed by the two main instruments on board the {\it INTEGRAL} observatory, IBIS and SPI \cite{Lau11, Jou12}. Both teams have demonstrated that the  400 -- 2000 keV emission in Cyg X-1 is polarized at a level of about $70 \%$, while only a $20 \%$ upper limit on the degree of polarization was obtained at lower energies. It has been suggested that the polarized emission is due to synchrotron emission from the compact jet. This result is important for our understanding of the accretion-ejection process.

\begin{figure}
\begin{center}
\begin{tabular}{c}
\begin{minipage}{0.5\linewidth}
\includegraphics[scale=0.32]{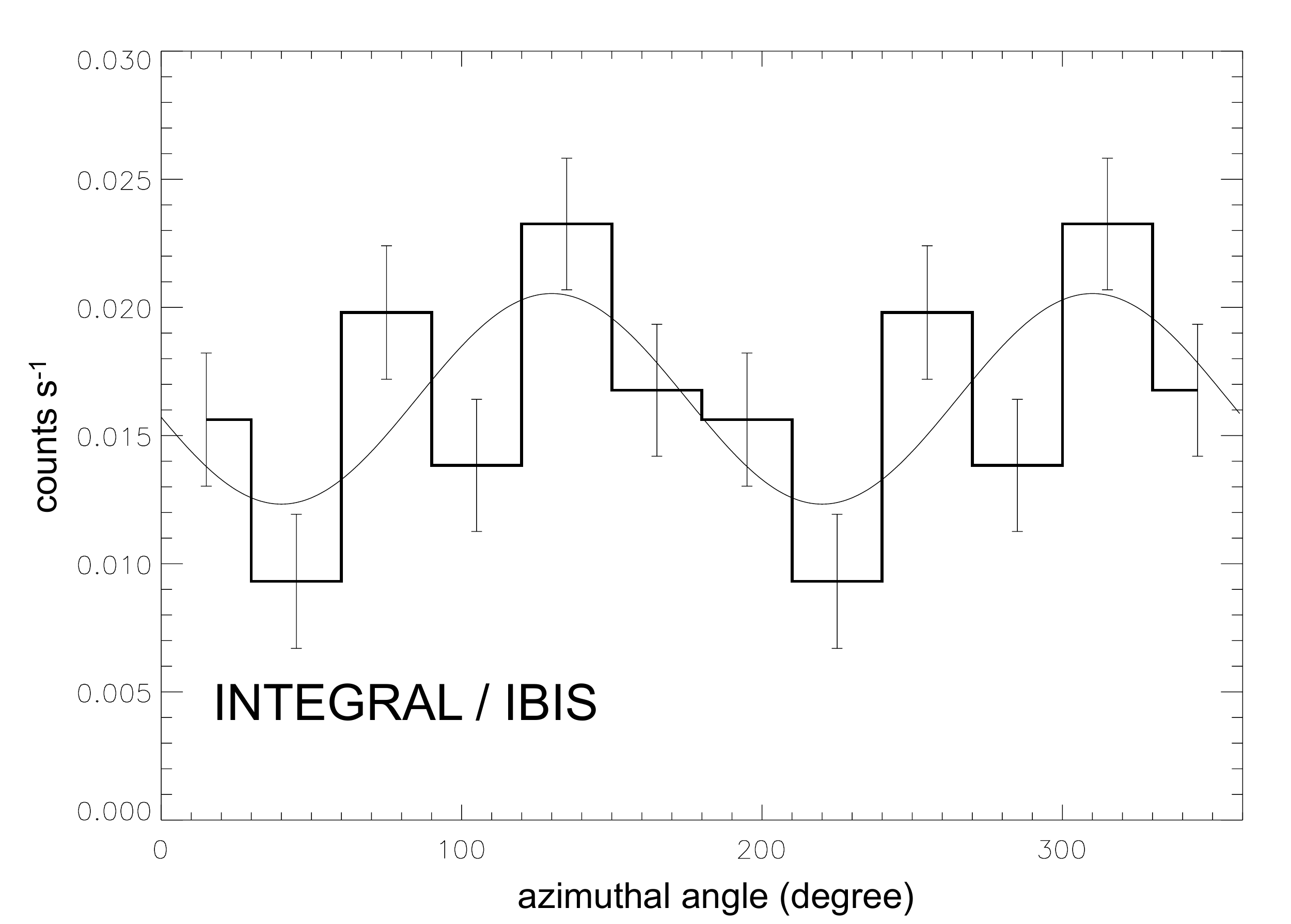}
\end{minipage}
\begin{minipage}{0.5\linewidth}
\includegraphics[scale=0.31]{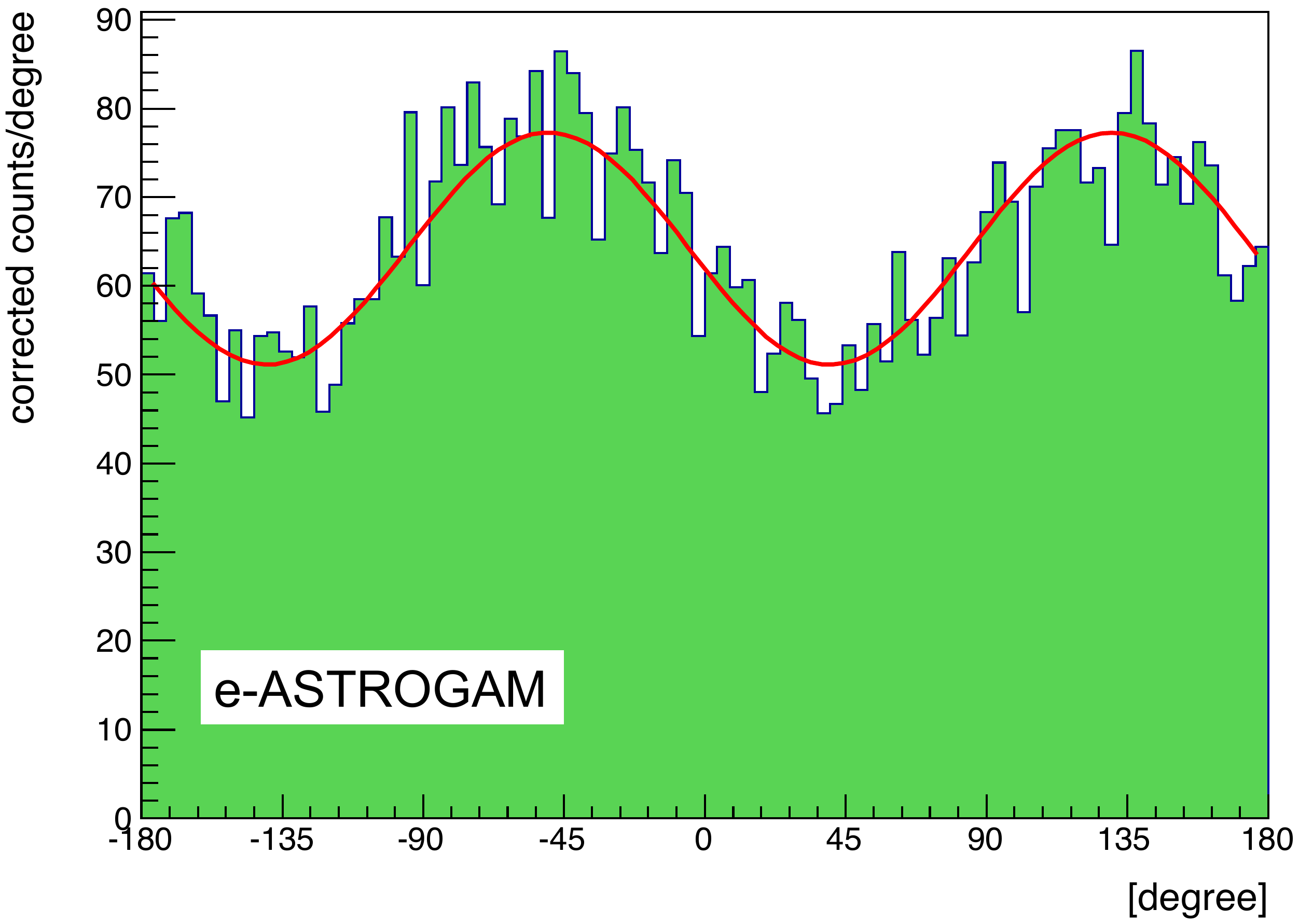}
\end{minipage}
\\
(a) \hspace{8.2cm} (b)
\end{tabular}
\end{center}
\caption 
{ \label{fig:CygX-1}
Similar to Fig.~\ref{fig:Crab}, but for the BHB Cygnus X-1 in the low hard state (LHS). (a) Polarization diagram obtained with {\it INTEGRAL}/IBIS in the 400 -- 2000 keV energy band by summing 2.05~Ms of data accumulated between May 2003 and May 2014 (adapted from Ref.~\citenum{rod15}). The measured polarization fraction and angle are ${\rm PF}=(75\pm32)$\% and ${\rm PA}=40.0^\circ \pm 14.3^\circ$, respectively. (b) Simulation of the e-ASTROGAM polarization signal in the same energy band (400 -- 2000 keV) for an effective duration of observation of Cyg X-1 in the hard state 100 times shorter than that accumulated by {\it INTEGRAL}/IBIS (i.e. $T_{\rm obs} = 2.05 \times 10^4$~s). The reconstructed polarization fraction and angle are ${\rm PF}=(75.0 \pm 4.7)$\% and ${\rm PA}=41.2^\circ \pm 1.9^\circ$.} 
\end{figure} 

Figure~\ref{fig:CygX-1} shows a comparison of a polarization diagram for Cygnus X-1 in the LHS obtained with {\it INTEGRAL}/IBIS summing 2.05~Ms of data\cite{rod15}, and a simulated polarigramme for e-ASTROGAM observation of Cyg X-1 in the LHS with 100 times shorter exposure  ($T_{\rm obs} = 2.05 \times 10^4$~s). For the simulations the energy spectrum of Cyg X-1 in the LHS measured by Rodriguez et al.\cite{rod15} was used, along with the central values obtained by these authors with {\it INTEGRAL}/IBIS for the polarization fraction (${\rm PF}=75.0$\%) and angle (${\rm PA}=40.0^\circ$) in the energy range $400$~--~$2000$~keV. The simulated result shows that e-ASTROGAM will be able to provide a detailed spectral and time-resolved gamma-ray view of the polarization properties of Cyg X-1, thus answering questions regarding the different emitting media in those sources (Comptonized corona vs. synchrotron-self Compton jets), while providing important clues concerning the composition, energetics and magnetic field of the jet. 

The excellent sensitivity of e-ASTROGAM in the MeV range will also give access to detailed studies of other, fainter, microquasars (e.g. GRS 1915+105, 1E1740.7-2942, GRS 1758-258, GRO J1655-40), whose polarized gamma-ray emission is not currently detectable. 

\section{Conclusions}
\label{sect:conclusions}  

The e-ASTROGAM gamma-ray telescope is based on an innovative design allowing measurements of the energy and 3D position of each interaction within the detectors with excellent spectral and spatial resolution, while minimizing passive material in the detector volume. With these features, e-ASTROGAM will be able to perform unprecedented polarization measurements in the MeV range using Compton interactions in the instrument. In the pair domain, a precise evaluation of the instrument polarimetric performance will require more detailed simulations. But preliminary estimates suggest that a 5$\sigma$ minimum detectable polarization of $\approx 20$~--~40\%  in the energy range from 10 to 100 MeV should be within reach for bright gamma-ray sources such as the Vela and Crab pulsars (see Ref.~\citenum{eas}).

e-ASTROGAM polarization sensitivity will complement the polarization studies at lower energies, such that of the IXPE mission, which is currently under construction at NASA\cite{wei16}. That mission, however, will be sensitive in the 1 -- 10 keV band, and together, the data from both missions will afford polarization measurements over an unprecedented spectral range.

Other projects are currently under study for the detection of linear polarization of high-energy cosmic gamma-rays above the $e^+$~--~$e^-$ pair production threshold, such as the GRAINE (Gamma-Ray Astro-Imager with Nuclear Emulsion\cite{oza16}), AdEPT (Advanced Energetic Pair Telescope\cite{hun14}), and HARPO (Hermetic ARgon POlarimeter\cite{gro16}) experiments.

The polarimetry capabilities of e-ASTROGAM will be crucial for a variety of investigations. Thus, it will bring a new diagnostic of gamma-ray emission models in all classes of AGN, with the potential to clearly identify the presence of hadrons in jets of BL Lac blazars, which are considered as a possible source of ultra-high-energy cosmic rays. In GRBs, polarization information can place strong new observational constraints on the nature of the ultrarelativistic outflow, its geometry and magnetization, as well as on the high-energy radiation mechanisms; measurements of GRB polarization also have the potential to provide constraining limits on violation of Lorentz invariance. e-ASTROGAM polarimetric studies of gamma-ray emission from compact objects will help to understand the new type of particle acceleration identified in the Crab nebula\cite{tav11,tav13}. In microquasars, e-ASTROGAM's sensitivity to polarization will make it possible to probe in detail the transition between accretion disks and relativistic jets, and will provide new insight into the composition, energetics and magnetic fields of these jets. For all of these studies, simulations of the e-ASTROGAM response to polarized photons show that the proposed space mission will be able to exploit, for the first time, the information encoded in the polarization of many types of cosmic gamma-ray sources. 

\acknowledgments 
The research leading to these results has received funding from the European Union's Horizon 2020 Programme under the AHEAD project (grant agreement n. 654215).

\bibliography{report}   
\bibliographystyle{spiejour}   

\end{spacing}
\end{document}